\documentclass[twocolumn]{bmcart}

\usepackage{amsthm,amsmath,amsfonts}
\usepackage[numbers]{natbib}
\usepackage[utf8]{inputenc} 
\usepackage{makecell}
\usepackage{siunitx}
\usepackage{graphicx}
\usepackage{soul}
\usepackage{multirow}

\startlocaldefs
\endlocaldefs

\begin{document}

\begin{frontmatter}

\begin{fmbox}
\dochead{Review}


\title{Automated Audio Captioning: An Overview of Recent Progress and New Challenges}


\author[
  addressref={aff1},                   
corref={aff1},                       
  email={x.mei@surrey.ac.uk}   
]{\fnm{Xinhao} \snm{Mei}}
\author[
  addressref={aff1},                   
  email={xubo.liu@surrey.ac.uk}   
]{\fnm{Xubo} \snm{Liu}}
\author[
  addressref={aff1},
  email={m.plumbley@surrey.ac.uk}
]{\inits{D.}\fnm{Mark} \snm{D. Plumbley}}
\author[
  addressref={aff1},
  email={w.wang@surrey.ac.uk}
]{\fnm{Wenwu} \snm{Wang}}


\address[id=aff1]{
  \orgdiv{Centre for Vision, Speech and Signal Processing (CVSSP)},             
  \orgname{University of Surrey},          
  \city{Guildford},                              
  \cny{UK}                                    
}





\begin{abstractbox}

\begin{abstract} 
Automated audio captioning is a cross-modal translation task that aims to generate natural language descriptions for given audio clips. This task has received increasing attention with the release of freely available datasets in recent years. The problem has been addressed predominantly with deep learning techniques. Numerous approaches have been proposed, such as investigating different neural network architectures, exploiting auxiliary information such as keywords or sentence information to guide caption generation, and employing different training strategies, which have greatly facilitated the development of this field. In this paper, we present a comprehensive review of the published contributions in automated audio captioning, from a variety of existing approaches to evaluation metrics and datasets. We also discuss open challenges and envisage possible future research directions.

\end{abstract}




\end{abstractbox}

\end{fmbox} 

\end{frontmatter}



\section{Introduction}
\label{sec1:intro}
Sound is ubiquitous in our daily lives. It carries a wealth of information about the environment, from sound scenes to individual events happening around us. For most people, the ability to perceive and understand the everyday sounds around us is taken for granted. However, mining helpful information from sounds is a challenging task for machines. With the development of machine learning, the field of machine listening has attracted increasing attention, with significant progress made in recent years, in areas such as audio tagging (AT) \cite{audioset, Xu2017unsupervised, kong2019weakly, Wang2020modelling, xu2017ijcnn}, sound event detection (SED) \cite{kong2019sed, kong2020sed_weakly, mesaros2021sound} and acoustic scene classification (ASC) \cite{barchiesi2015acoustic, Wang2021sa++}. However, in these areas, the focus has been mostly on identifying acoustic scenes or events in an audio clip, rather than considering relationships between the audio events and acoustic scenes. 

Automated audio captioning (AAC) aims at describing the content of an audio clip using natural language, which is a cross-modal translation task at the intersection of audio signal processing and natural language processing (NLP) \cite{drossos2017ac_1}. Compared with automatic speech recognition (ASR), audio captioning focuses only on the environmental sounds and ignores the voice content that may be present in an audio clip. Compared with other popular audio-related tasks such as AT, SED, and ASC, audio captioning requires not only determining what audio events are present in the audio clip, but also describing these audio events using natural language, which allows the relationships between the audio events and the content of the audio clip to be summarized.
An example caption may be ``a person was walking on a sidewalk adjacent to a school where children were playing on the playground''\footnote{This caption is from the Clotho dataset.} which describes the scenes and sound events given an audio clip. Generally speaking, audio captions are one-sentence descriptions of the predominant audio events and audio scenes occurring in the audio clips, where the detailed information may be included, such as the spatial-temporal relationships between the audio events and scenes, and the physical properties of sound objects and the acoustic environment.

Audio captioning has practical potential for various applications such as helping the hearing-impaired to understand environmental sounds, and analyzing sounds for video-based security surveillance systems. In addition, audio captioning can be used for multimedia retrieval \cite{koepke2022audioretrieval, mei2022metric} in areas including education, film production, and web searching.

\begin{figure*}[ht]
    \centering
    \includegraphics{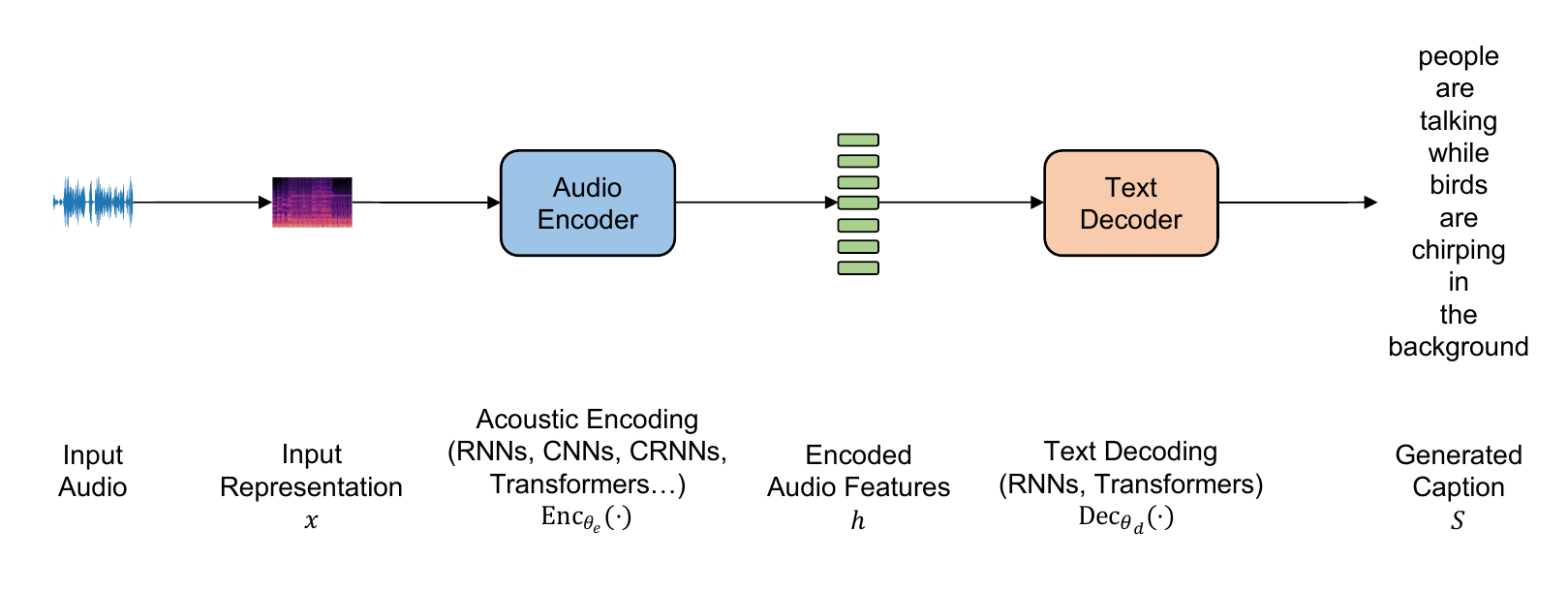}
    \caption{Overview of an encoder-decoder-based AAC system, where the input is the waveform of an audio clip and the output is a natural language sentence describing the content of the input audio clip.}
    \label{fig1:encoder_decoder}
\end{figure*}

Unlike image and video captioning, which have been widely studied for almost a decade, audio captioning is a relatively new task that has been studied since 2017 \cite{drossos2017ac_1}. In the past three years, this field has received increasing attention due to the release of freely available datasets and the organisation of a task in DCASE\footnote{http://dcase.community/} Challenges from 2020 to 2022. A number of papers about audio captioning have been published, with deep learning being a popular method. Specifically, the encoder-decoder framework \cite{sutsjever2014enc_dec} has been adopted as a standard recipe for solving this cross-modal translation task. In this method, the encoder extracts audio features from the input audio clips, and the decoder generates captions based on the extracted audio features. 
Analyzing audio largely depends on obtaining robust audio features. Different kinds of neural networks, such as Recurrent Neural Networks (RNNs) \cite{rumelhart1986rnn}, Convolutional Neural Networks (CNNs) \cite{lecun2015deep}, and Transformers \cite{vaswani2017attention}, have been used as the encoders to learn feature representations. For the decoder, RNNs and Transformers are usually employed, inspired by works in NLP. In addition to the encoder-decoder framework, auxiliary information such as keywords or sentence information \cite{koizumi2020keywords, xu2021audiocaption_car}, attention-based approaches \cite{kim2019audiocaps, drossos2017ac_1} and different training strategies \cite{xu2021_sjtu, Berg2021continual, Liu2021cl4ac} have been proposed to improve the performance of captioning systems. However, there is still a large gap between achieved results and human level performance \cite{kim2019audiocaps}.

To the best of our knowledge, no survey papers on audio captioning have been published so far. In this paper, we aim to provide a comprehensive overview of audio captioning with the hope of stimulating novel research ideas. Articles published up to April 2022 in the literature are considered in our survey. The encoder-decoder framework has been a standard recipe for AAC systems, therefore, we develop a taxonomy of acoustic encoding and text decoding approaches.

This paper is organized as follows. Section~\ref{sec2:pre} introduces the preliminaries of audio captioning. In Section~\ref{sec2:acoustic} and Section~\ref{sec:text}, we discuss acoustic encoding and text decoding approaches respectively. Auxiliary information is discussed in Section~\ref{sec:auxiliary}. We discuss training strategies adopted in the literature in Section~\ref{sec:training}. Furthermore, we review popular evaluation metrics and main datasets in Section~\ref{sec:eval_metrics} and Section~\ref{sec:datasets}, respectively. Finally, we discuss some open challenges and future research directions in Section~\ref{sec:challenges} and briefly conclude this paper in Section~\ref{sec:conclu}.

\section{Preliminaries of audio captioning}
\label{sec2:pre}
Existing methods for audio captioning are built predominantly on an encoder-decoder architecture where the captions are generated in an auto-regressive manner using deep learning techniques. We, therefore, take the popular encoder-decoder architecture as an example to introduce the preliminaries of an audio captioning system. Figure \ref{fig1:encoder_decoder} shows the pipeline of an AAC system based on the encoder-decoder architecture.

Suppose we have a raw waveform of an input audio clip. Human-engineered features are usually extracted from the waveform as input representations for the audio encoder. Assume here that mel-spectrogram is used as the input representation, denoted by $x$, with a shape of $\mathbb R^{T\times F}$, where $T$ is the number of time frames and $F$ is the number of mel bins. The audio encoder takes the mel-spectrogram $x$ as input and produces the encoded audio features $h$, which could be a single vector of shape $\mathbb R^C$, or a vector sequence of shape $\mathbb R^{T'\times C}$ where $C$ is the dimension of the audio feature and $T'$ is the number of feature vectors, depending on the type of the encoder and the pooling method used for learning the encoded audio features. This process can be formulated as follows:
\begin{equation}
    \label{eqn:encoder}
  h = {\rm Enc}_{\theta_e}(x)
\end{equation}
where $\theta_e$ are the model parameters of the encoder (Enc). More discussions about how the features are learned are given in Section \ref{sec2:acoustic}.

After getting the encoded audio feature $h$, the decoder generates a sentence $S=\{w_1,..., w_N\}$, where $w_n$ is a word and $N$ is the number of words in the sentence. The decoding process can be formulated as follows:
\begin{equation}
  \label{eqn:decoder}
  S = {\rm Dec}_{\theta_d}(h)
\end{equation}
where $\theta_d$ are the model parameters of the decoder (Dec). Typically, the sentence is generated from the left (i.e. the first word) to the right (i.e. the final word) in an auto-regressive manner. That is, at time step $t$, the decoder predicts a posterior probability over the vocabulary, given the encoded audio feature $h$, a start token $w_0$, and previously generated words $w_1$ to $w_{t-1}$. 
Mathematically,
\begin{equation}
  \label{eqn:decoder_pre}
  p(w_t|h,w_0,...,w_{t-1}) = {\rm Dec}_{\theta_d}(h,w_0, ...,w_{t-1}),
\end{equation}
where $w_0$ is a starting word of the sentence. After obtaining the word probability $p(w_t|h,w_0,...,w_{t-1})$, the word $w_t$ can be sampled by different decoding methods, such as greedy decoding or beam decoding \cite{graves2012sequence}. The generation process is terminated when a stop token is generated or a maximum number of generation steps is reached.

\section{Acoustic encoding}
\label{sec2:acoustic}
Analyzing the content of an audio clip largely depends on obtaining an effective feature representation for it, which is the aim of the encoder in an AAC system. The time domain waveforms are lengthy 1-D signals and it is challenging for machines to directly identify sound events or sound scenes from raw waveforms \cite{virtanen2018computational}. Current popular approaches for acoustic encoding consist of two steps, first extracting input representations, which are often hand-crafted features, such as spectrograms from the audio clip, and then passing them into a neural network to learn encoded compact audio features. In this section, we first discuss popular hand-crafted features used in literature, then audio encoding approaches, focusing on those based on deep neural networks.

\subsection{Hand-crafted features}
\label{ssec2:ac_feature}
It is challenging for machines to directly understand an audio clip from its time domain representation. Hand-crafted features, inspired by the human auditory system have been widely used as sound representations for years \cite{virtanen2018computational}. In deep learning methods, these hand-crafted features are used as input representations to the neural networks to obtain encoded audio features. Time-frequency representations, such as spectrograms, are probably the most popular ones. To obtain a spectrogram, an audio signal is first split into short frames of length at around \num{20}-\num{60} ms, as these short time segments can be regarded as quasi-stationary \cite{virtanen2018computational}. Each time frame is shifted with a fixed time step. Then a window function is applied at each frame to enforce continuity and avoid spectral leakage at the frame boundaries \cite{harris1978leakage}. The short time Fourier transform (STFT) is calculated for each time frame to get the spectrogram, a 2-D representation whose horizontal axis is time and vertical axis is frequency, the value at each point of the spectrogram represents the magnitude at a specific time and frequency. Inspired by the selectivity of human auditory system to different frequencies, the frequency axis of a spectrogram may be converted to different scales, resulting in representations such as mel-spectrogram and log mel-spectrogram \cite{virtanen2018computational}. The log mel-spectrogram generally leads to better performance when compared with other input representations in deep learning based methods for audio-related tasks \cite{kong2020panns, hershey2017vggish, wu2019audiocaption_hospital}, therefore, it is mainly used as the input representation in the literature. In addition, mel-frequency cepstral coefficients (MFCCs) were used in some early works \cite{kim2019audiocaps, Ikawa2019ac_spec}. MFCCs are calculated by applying a discrete cosine transform (DCT) on log mel-spectrograms. Compared with time-frequency representations, MFCCs contain less information and are only able to estimate the global spectral shape of an audio clip \cite{virtanen2018computational}, thus MFCCs are rarely used in recent works.

\subsection{Neural networks}
\label{ssec2:ac_networks}

\begin{figure}[t!]
    \centering
    \includegraphics{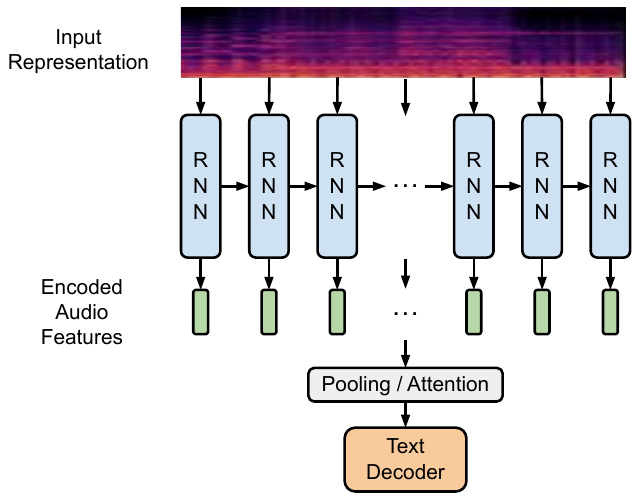}
    \caption{Diagram of an RNN audio encoder for acoustic encoding. The RNN encoder aims at modelling temporal relationships within the input representation. The encoded audio features usually have the same number of time frames as the input representation and interact with the decoder through a pooling or attention mechanism.}
    \label{fig:rnn_encoder}
\end{figure}

\subsubsection{RNNs}
RNNs are designed to process sequential data with variable lengths \cite{rumelhart1986rnn}. Audio is time series signal, therefore RNNs are naturally adopted as encoders in initial works \cite{drossos2017ac_1, Ikawa2019ac_spec}. In a simple recipe, an RNN is used to model temporal relationships between the inputs, and the hidden states of the last layer of the RNN are regarded as the audio feature sequences, which are then fed into the text decoder for caption generation. Figure~\ref{fig:rnn_encoder} shows the diagram of an RNN audio encoder. Drossos et al. \cite{drossos2017ac_1} utilized a three-layered bi-directional gated recurrent unit (GRU) network \cite{chung2014gru} as the encoder. Further, unlike using multi-layer RNNs, Xu et al. \cite{xu2021audiocaption_car} and Wu et al. \cite{wu2019audiocaption_hospital} used a single-layered uni-directional GRU network while Ikawa et al. \cite{Ikawa2019ac_spec} used a single-layered bi-directional long-short term memory (LSTM) network \cite{hochreiter1997lstm}. The encoded audio features output by RNNs usually have thousands of time steps, Nguyen et al. \cite{Nguyen2020temporalsub} argued that the length of the captions is significantly less than the length of the encoded audio features, making the captioning models difficult to learn the correspondence between words and audio features. They proposed a temporal sub-sampling method to sub-sample the learned features between the RNN layers, and showed that the temporal sub-sampling of audio features could be beneficial for audio captioning methods.

The main advantages of employing RNNs as encoders are their simplicity and their ability to process sequential data. However, using RNNs alone as the encoder is not found to give strong performance \cite{kim2019audiocaps}. The reason might be that inputs are usually long sequences, RNNs may not be able to effectively model long-range time dependencies. In addition, getting an global audio feature from long hidden states also leads to excessive compression of information, making it difficult for the language decoder to generate fine-grained descriptions. 

\begin{figure}[t!]
    \centering
    \includegraphics{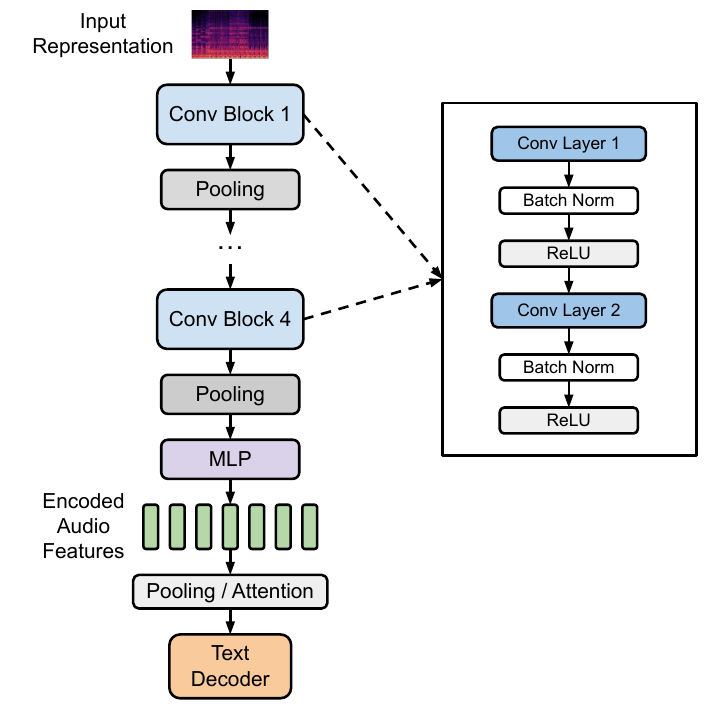}
    \caption{Diagram of a 10-layer CNN audio encoder. The input representation is first processed via four convolutional blocks and pooling layers, where each block consists of two convolutional layers. The feature maps output by the last convolutional block are then averaged along the frequency axis and fed into a two-layer multi-layer perceptron (MLP) to obtain encoded audio features.}
    \label{fig:cnn_encoder}
\end{figure}

\subsubsection{CNNs}
CNNs have been applied with great success to the field of computer vision (CV) \cite{lecun2015deep}. In recent years, CNNs have been adapted to audio-related tasks and show powerful ability in extracting robust audio patterns \cite{hershey2017vggish, kong2020panns}. Figure~\ref{fig:cnn_encoder} shows the diagram of a 10-layer CNN audio encoder that is popularly used in the literature \cite{chen2020ac_CNN, Mei2021ac_trans}.

Many CNN models pre-trained on large audio datasets have been published. Most works directly employ pre-trained CNN models as the audio encoder. VGG-like CNNs \cite{chen2020ac_CNN, Mei2021ac_trans} and ResNets \cite{Ye2021peking, Han2021netease, Perez-Castanos2020ac_listen} are popular choices as these networks perform well on audio-related tasks such as audio tagging and sound event detection \cite{kong2020panns}. In these works, CNNs treat the input spectrograms as 1-channel images, and model local dependencies within the spectrograms. Moreover, 1-D CNN is also incorporated to exploit temporal patterns. For example, Eren et al. \cite{eren2020acfake} and Han et al. \cite{Han2021netease} used Wavegram-Logmel-CNN adapted from pre-trained audio neural networks (PANNs) \cite{kong2020panns}. The Wavegram-Logmel-CNN takes both raw waveform and spectrogram as inputs, which are processed using 1-D convolution and 2-D convolution, respectively. The outputs of 1-D convolutional layers and those of 2-D convolutional layers are combined in deep layers. Tran et al. \cite{tran2020wavetransformer} also proposed to utilise 1-D and 2-D convolutions for extracting temporal and time-frequency information. However, they only used spectrogram as input and reshape it for 1-D convolution. In summary, the use of 1-D convolution requires increased computation overhead, but offers only small performance improvement. The output feature maps of the convolutional blocks are generally in three dimensions, time, frequency and channels. To obtain encoded audio features, some methods use a global pooling along the time and frequency axis to obtain fixed-sized features \cite{eren2020acfake}, while others keep the time axis and apply pooling operation along the frequency axis to get a feature sequence \cite{Mei2021ac_trans, Ye2021peking}.

In summary, CNNs outperform RNNs, and are now the dominant approach for audio encoding. The main advantages of CNNs are that they are invariant to time shift and good at modeling local dependencies within the spectrograms. However, CNNs have limited receptive fields and modeling long-range time dependencies for long audio signals needs a deep CNN. 

\subsubsection{CRNNs}
Motivated by the demand for modelling the local and long-range dependencies simultaneously, convolutional recurrent neural networks (CRNNs) \cite{shi2016crnn}, a combination of CNNs and RNNs, have also been applied as audio encoders. In a CRNN, RNN layers are introduced after the CNN layers to model the temporal relationship between extracted CNN features. Kim et al. \cite{kim2019audiocaps} proposed a top-down multi-scale encoder where the features are extracted from two layers of the VGGish network \cite{hershey2017vggish}, that is, a fully connected layer for extracting the high-level semantic features and a convolutional layer for extracting the mid-level features. Those features are then encoded by a two-layer bi-directional LSTM network where the semantic features are injected in the second layer. Takeuchi et al. \cite{Takeuchi2020ntt_first} and Xu et al. \cite{xu2020crnn_sjtu} both adopted a similar CRNN encoder without using multi-level features. Xu et al. \cite{xu2021invest_cnn_crnn} compared CNN and CRNN encoders, and showed that a CRNN encoder outperformed a CNN encoder when the encoders are trained from scratch but the CRNN encoder brought little improvement when pre-training was applied. In summary, CRNNs need more computation than CNNs but offer limited improvement \cite{xu2021invest_cnn_crnn}.

\begin{figure}[t!]
    \centering
    \includegraphics{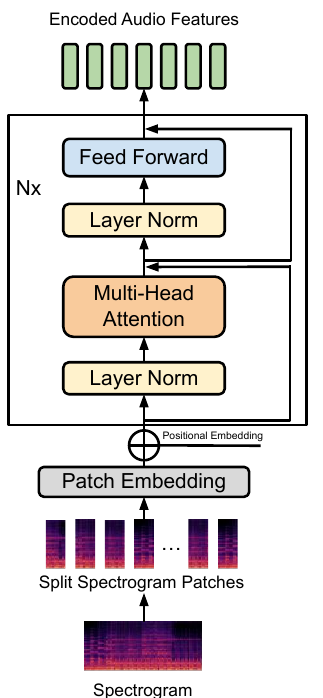}
    \caption{Diagram of the Transformer-based audio encoder. The input spectrogram is first split into small patches. These patches are then projected into 1-D embeddings through a linear layer, where a positional embedding is further added to each patch embedding to capture position information. The resulting embeddings are then fed into the Transformer blocks to obtain the encoded audio features.}
    \label{fig:trans_encoder}
\end{figure}

\subsubsection{Other approaches}
Transformers and their variants that are built on self-attention mechanism have been probably the most popular models in the fields of NLP and CV since 2017 \cite{vaswani2017attention, dosovitskiy2020image, touvron2021training}. Self-attention based encoders are also employed in recent works in audio captioning. Koizumi et al. \cite{koizumi2020keywords} introduce a self-attention block after CNN layers in the encoder to learn the temporal relationship between CNN features. Mei et al. \cite{Mei2021ACT} proposed Audio Captioning Transformer (ACT), where the encoder is a convolution-free Transformer that directly models the relationships between the patches of the spectrogram. Figure~\ref{fig:trans_encoder} shows the diagram of the Transformer-based audio encoder in ACT. More details about Transformer and self-attention will be introduced in Section~\ref{ssec:trans}. ACT shows comparable performance with CNN-based methods while it may need more data for pre-training to obtain good performance. In addition to simply adding self-attention layers after convolutional layers, convolution and self-attention can be combined as in \cite{Narisetty2021ac_conformer} by leveraging a convolution-augmented Transformer (Conformer) \cite{gulati2020conformer} to take advantage of their respective strengths. However, the Conformer encoder did not outperform the CNN encoders. The reason might be that they did not pre-train the Conformer encoder on a large-scale audio dataset.

In summary, various neural network architectures have been investigated as the audio encoder in order to obtain robust audio representations. CNNs are probably the most popular audio encoders and have achieved state-of-the-art performance. Early works adopted RNNs as encoders, but the trend has shifted from RNNs to CNNs. Recently, novel Transformer-based architectures have received increasing attention and have shown competitive performance in learning robust audio features as compared with the CNN encoders, however, they typically require more data for training to achieve similar performance, in comparison to the CNN networks \cite{Mei2021ACT}.

\section{Text decoding}
\label{sec:text}
The aim of the text decoder is to generate a caption given the audio features from the encoder. Existing works adopt an auto-regressive method for text generation, where each word in the caption is predicted based on the condition of the audio features extracted by the encoder and previously predicted words by the decoder. In addition to the main decoder block, a word embedding layer is used before the main decoder block to embed each input word into a fixed-dimension vector, so that discrete words can be processed by the network. In this section, we first introduce popular methods for obtaining word embeddings and then discuss main approaches for text decoding.

\subsection{Word embeddings}
\label{ssec:word}
A simple method to obtain word vectors is to represent each word as a one-hot vector, in which the element whose position corresponds to the index of the word in the vocabulary is set to one, while the remaining elements are set to zeros. This is called one-hot encoding. If the vocabulary is large, the dimension of the one-hot vector can be high. Hence, this method may suffer from the curse of dimensionality and the loss of semantic information \cite{jurafsky2009slp2}. Word embedding methods have become popular in recent years. Word embeddings are vectors of fixed-dimension, obtained by neural networks trained on large-scale text corpora. Semantically similar words are close to each other in the embedding space, while dissimilar words are far away from each other \cite{mikolov2013word2vec}. Examples of pre-trained word embeddings include Word2Vec \cite{mikolov2013word2vec}, GloVe \cite{pennington2014glove} and fastText \cite{mikolov2018fasttext}, which are widely used in existing audio captioning works \cite{kim2019audiocaps, chen2020ac_CNN, xu2020crnn_sjtu, koizumi2020keywords, eren2020acfake}. 

Recently, large-scale pre-trained Transform-based language models \cite{devlin2019bert, radford2019gpt2} have shown powerful ability in language modelling thanks to the use of the self-attention mechanism. Weck et al. \cite{Weck2021ac_offtheshelf} employed BERT \cite{devlin2019bert} to obtain word embeddings. They compared the effect of different pre-trained models on obtaining the word embeddings, and found that the BERT model leads to the best performance in obtaining word embeddings, while other models, such as Word2Vec and GloVe, provide slight improvement as compared to randomly initialized word embeddings.

In summary, word embeddings are semantic vector representations of words. They are generally stored in a matrix with the shape of $\mathbb R^{V\times d}$, where $V$ is the size of the vocabulary and $d$ is the dimension of the word vector. They can be retrieved using the indices of the words in the vocabulary.

\subsection{Neural networks}

\begin{figure}[t!]
    \centering
    \includegraphics{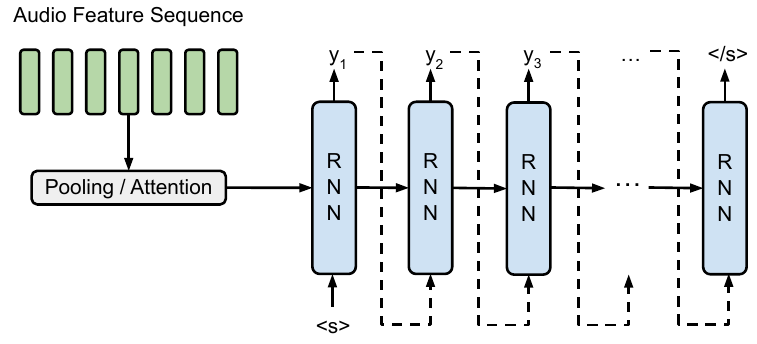}
    \caption{Diagram of an RNN-based language model. The RNN decoder generates the sentence from the left (i.e. the first word) to the right (i.e. the final word) in an auto-regressive manner, given the audio feature sequence generated from the encoder and previously generated words by the decoder. A start token `\textless s\textgreater' is fed into the RNN at the first time step to start the generation, while the generation process is terminated when a stop token `\textless/s\textgreater' is generated.}
    \label{fig:rnn_lan}
\end{figure}

\subsubsection{RNNs}
Sentences are also sequential data composed by discrete words, thus RNNs are popularly employed as the language decoder. Figure~\ref{fig:rnn_lan} shows a diagram of RNN-based language decoder. 
At each time step, the hidden state of the RNN is projected into a probability distribution along the vocabulary through a linear layer with  a softmax activation function, and a word can be predicted accordingly.

Drossos et al. \cite{drossos2017ac_1} proposed a 2-layer GRU network as the decoder in their initial work. Many subsequent works have adopted single-layer RNNs, either GRU or LSTM networks \cite{Ikawa2019ac_spec, kim2019audiocaps, Nguyen2020temporalsub, Cakir2020ac_multitask, Ye2021peking}. The main differences among these works are on how the audio features generated by the encoder are fused with the decoder. In a simple recipe, a global audio feature representation is obtained by applying mean pooling on the audio feature sequence extracted by the encoder, which is then used as the initial hidden state of the RNN decoder \cite{wu2019audiocaption_hospital, xu2021audiocaption_car} or is injected to the RNN decoder at each time step \cite{Nguyen2020temporalsub, eren2020acfake}. This simple mean pooling method for getting a global audio representation is widely used in audio tagging task to detect what audio events are present in the whole audio clip \cite{kong2020panns}. However, this method does not consider the relationships between audio features, and thus, it is unable to capture the fine-grained information about audio events. These fine-grained information could be important for caption generation. Attention mechanism has been employed to overcome this problem \cite{drossos2017ac_1}. When generating a word at each time step, the RNN decoder can attend to the whole audio feature sequence and place more weights on the informative audio features. Thus, the global audio representation at each time step is a different combination of the whole audio feature sequence. In addition, to exploit previously generated words, Ye et al. \cite{Ye2021peking} introduced another attention module to attend to previously generated words at each time step. 

In summary, RNNs with attention show reasonable performance in audio captioning and are widely used \cite{xu2020crnn_sjtu, Ye2021peking}. The main disadvantage of RNNs is that they may be struggling to capture long-range dependencies between the generated words. Fortunately, audio captions are usually short in length, thus RNN decoders do not need to model very long-range dependencies.

\begin{figure}[t!]
    \centering
    \includegraphics{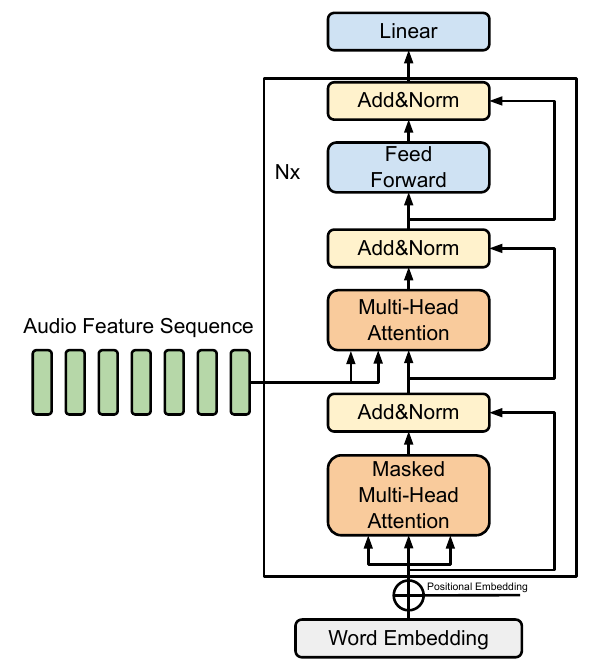}
    \caption{Diagram of a Transformer-based language model. When generating a word at each time step, the masked multi-head attention module attends to the previously generated words to exploit contextual information. The output of the masked self-attention module is then fused with the audio feature sequence from the encoder in the cross-attention module.}
    \label{fig:transformer_lan}
\end{figure}

\subsubsection{Transformers}
\label{ssec:trans}
Since Vaswani et al. \cite{vaswani2017attention} proposed the Transformer network in 2017, the self-attention mechanism, which is the core of Transformer, has quickly become the basic building block in large language models. Transformer-based models such as BERT \cite{devlin2019bert}, GPT \cite{radford2019gpt2} and BART \cite{lewis2020bart} outperform RNNs in language modelling and dominate most of the tasks in the field of NLP. Transformer-based models have also been employed in audio captioning works recently and achieved state-of-the-art performance.

Transformers are built on the self-attention mechanism. The self-attention module takes a sequence of $N$ inputs and return $N$ outputs, allowing each input directly interacts with others within the input sequence and finds out which they should pay more attention to. This makes it easier to model long range and global dependencies within the sequence, as compared with RNNs. Concretely, given an input sequence, the self-attention module first transforms the inputs into three representations, query vectors $Q$, key vectors $K$, and value vectors $V$ by three learnable matrices. For each input, a scaled dot-product is first calculated between its query with respect to all keys to obtain the attention weights. After that, the attention weights are first converted to probabilities by a softmax function and then multiplied with each value and summed together to get the output. This can be formulated as:
\begin{equation}
  \label{eqn:attention}
  {\rm Attn}(Q,K,V) = {\rm Softmax}(\frac{QK^T} {\sqrt{d_k}})V.
\end{equation}
where $d_k$ is a scaling factor. In addition, this computation can be parallelized for all inputs through matrix multiplication, the training of Transformer could be more efficient than that of RNNs. The Transformer decoder is generally employed as the language decoder and has a stack of blocks, each of which consists of a masked self-attention module, a cross-attention module and a feed-forward layer module. Figure~\ref{fig:transformer_lan} shows a diagram of Transformer-based language decoder. The audio feature sequence obtained from the encoder interacts with the decoder through the cross-attention module, where $K$ and $V$ are derived from the audio features, while $Q$ is obtained from the output of the masked self-attention module. Through these two attention modules, each word in the sequence can attend to the previously generated words and all the audio features, which may facilitate the model to capture the temporal relationships between audio events.

Due to the limited data available in audio captioning, many works use shallow Transformer decoders \cite{chen2020ac_CNN, Mei2021ac_trans, Han2021netease, Mei2021ACT}, usually only two blocks, unlike in NLP tasks where very deep Transformers are often used \cite{devlin2019bert, lewis2020bart}. Some modifications to the standard Transformer architecture have also been investigated. For example, Xiao et al. \cite{xiao2022local} introduced an attention-free Transformer decoder to reduce computation overhead, which they  claimed could better capture local information within audio features.

In summary, Transformer-based decoders show state-of-the-art performance in audio captioning, and are computationally more efficient from RNN-based decoders during training.

\section{Auxiliary information}
\label{sec:auxiliary}
In addition to the standard encoder-decoder architecture, researchers have investigated the use of auxiliary information such as keywords or sentence information to guide caption generation. In this section, we discuss the auxiliary information used in the literature. 

Audio signals have variable lengths and sound events can occur over arbitrary time frames, making direct mapping audio signals to sentences challenging. Furthermore, each sound event can be described with different words, which may lead to the word-selection indeterminacy problem \cite{koizumi2020keywords}. To improve caption generation, keywords are widely employed. To obtain keywords, Kim et al. \cite{kim2019audiocaps} retrieve the most similar training audio clip from AudioSet, the largest dataset for audio tagging available so far, and convert audio tagging labels of the retrieved audio clip into keywords. They then align these keywords with the audio features via an attention module and feed the output into the decoder. Some datasets may not have corresponding label information for each audio clip, and in this case, researchers first extract keywords or tags from human-annotated captions according to some rules such as frequency of the words and part-of-speech of the words \cite{chen2020ac_CNN, koizumi2020keywords, Cakir2020ac_multitask, Han2021netease, mei2022_t6a}. Different methods were investigated to make use of the keywords. Cakir et al. \cite{Cakir2020ac_multitask} introduce a keyword decoder to estimate keywords of an audio clip, and jointly train the keyword decoder with the audio captioning model. Chen et al. \cite{chen2020ac_CNN} extract keywords from captions, and pre-train the audio encoder with an audio tagging task to enhance the ability of the encoder to learn robust audio patterns. Koizumi et al. \cite{koizumi2020keywords} employ a keyword estimation branch after the encoder, combining the keywords with audio features before passing them to the language decoder. Ye et al. \cite{Ye2021peking} utilize multi-scale features extracted by a CNN encoder for keyword prediction. However, some researchers found that keywords might not really improve the system performance in some situations. Takeuchi et al. \cite{Takeuchi2020ntt_first} found that keywords may not work well when the model was trained from scratch. Ye et al. \cite{Ye2021peking} claimed their model did not converge when only using keywords information. The accuracy of the keywords could be a bottleneck as wrong keywords might adversely impact on the captioning performance.

Sentence information has also been investigated. Ikawa et al. \cite{Ikawa2019ac_spec} introduce a `specificity' term to measure the output text based on the amount of information it carries. The model is trained to generate captions whose `specificity' is close to ground truth captions. Similarly, Xu et al. \cite{xu2021audiocaption_car} introduce a sentence loss to generate captions closer to their ground truths in the sentence embedding space, employing a pre-trained language model BERT \cite{devlin2019bert} to get the sentence embeddings.

Although different auxiliary information has been used to improve the caption generation process, these methods have not brought significant improvements and they may not work well for all datasets. In the DCASE challenge 2021, most teams still used the standard encoder-decoder model without using auxiliary information, and still achieved promising results \cite{xu2021_sjtu, Mei2021ac_trans}. How to improve the AAC system with the auxiliary information still needs more investigation.

\section{Training strategies}
\label{sec:training}
Supervised training with a cross-entropy (CE) loss is a standard recipe for training an audio captioning model. The main drawback of this setting is that it may cause `exposure bias' due to the discrepancy between training and testing \cite{rennie2017exposurebias}. Reinforcement learning has been introduced to solve this problem and directly optimize evaluation metrics. In addition, transfer learning has been widely used to overcome the data scarcity problem. In this section, we discuss the popular training strategies used in the literature.

\subsection{Cross-entropy training}
The cross-entropy loss with maximum likelihood estimation (MLE) is widely used for training audio captioning models. During training, this approach adopts a `teacher-forcing' strategy \cite{rennie2017exposurebias}. That is, the objective of training is to minimize the negative log-likelihood (equivalent to maximizing the log-likelihood) of current ground truth word given previous ground truth words at each time step. The cross-entropy loss can be formulated as follows:
\begin{equation}
  \label{eqn:ce_loss}
  \mathcal L_{\rm CE}(\theta)= - \frac{1}{T} \sum_{t=1}^T \log{p(y_t|y_{1:t-1}, x, \theta)}
\end{equation}
where $y_t$ is the ground truth word at time step $t$, $T$ is the length of the ground truth caption, $x$ is the input audio clip, and $\theta$ are the parameters of the audio captioning model. 

Models trained via the cross-entropy loss can generate syntactically correct sentences and achieve high scores in terms of the evaluation metrics \cite{Mei2021ac_trans}. However, there are also some disadvantages. First, the `teacher forcing' strategy brings the problem known as `exposure bias' \cite{rennie2017exposurebias}, that is, each word to be predicted is conditioned on previous ground-truth words in the training stage, while it is conditioned on previous output words in the test stage. This discrepancy leads to error accumulation during text generation in the test stage. Second, models tend to generate generic and simple captions even though each audio clip has multiple diverse human-annotated captions in the training set \cite{mei2021diverse}. This is because the MLE training tends to encourage the use of highly frequent words appearing in the ground truth captions.

\subsection{Reinforcement learning}
Xu et al. \cite{xu2020crnn_sjtu} employ a reinforcement learning approach to solve the `exposure bias' problem and directly optimize the non-differentiable evaluation metrics. In a reinforcement learning setting, the captioning model is regarded as an agent and a policy is determined by the model's parameters. The agent executes an action at each time step to sample a word according to the policy. Once a sentence is generated, the agent receives a reward of the generated sentence. The goal of training is to optimize the model to maximize the expected reward that could be any evaluation metrics. This can be formalized as minimizing negative expected reward:
\begin{equation}
  \label{eqn:rl_loss}
  \mathcal L_{\rm RL}(\theta)= - E_{w^s \sim p_\theta}[r(w^s)],
\end{equation}
where $w^s$ is a sampled caption from the model, $r$ is the reward of the sampled caption and $\theta$ are the model parameters. The caption can be sampled via Monte-Carlo sampling \cite{sutton2018reinforcement}, however, it is computationally expensive. Another computationally efficient method, self-critical sequence training (SCST) \cite{rennie2017exposurebias} is generally employed. SCST employs the reward of a sentence sampled by greedy search as baseline thus avoids learning an estimate of expected future rewards. The expected gradient with respect to a single sample caption $w^s \sim p_\theta$ can be approximated as:
\begin{equation}
  \label{eqn:rl_gradient}
  \nabla_\theta \mathcal L_{\rm RL}(\theta) \approx -(r(w^s) - r(\hat{w}))\nabla_\theta \log p_\theta (w^s),
\end{equation}
where $r(\hat{w})$ is the reward of a caption generated by the current model using a greedy search.

Reinforcement learning could substantially improve the scores of the evaluation metrics, although it is used to optimize only one metric. However, Mei et al. \cite{Mei2021ac_trans} found that reinforcement learning may impact adversely on the quality of generated captions by introducing repetitive words and incorrect syntax. This also implies that existing evaluation metrics may not correlate well with human judgements.

\subsection{Transfer learning}
Availability of audio captioning datasets is limited due to the challenging and time consuming process in data collection and annotation \cite{drossos2020clotho,Martin2021databias}. To overcome the data scarcity problem, transfer learning is widely adopted. In the encoder of the captioning system, pre-trained audio neural networks such as VGGish \cite{hershey2017vggish} and PANNs \cite{kong2020panns}, are widely used to initialize the parameters of encoders \cite{Mei2021ac_trans, Ye2021peking, Han2021netease, koh2022ac_latent, wu2022wav2clip}. Xu et al. \cite{xu2021invest_cnn_crnn} investigated the impact of pre-training on audio captioning performance. They show that audio encoders pre-trained with an audio tagging task give the best performance. Weck et al. \cite{Weck2021ac_offtheshelf} compare four off-the-shelf audio networks. In all the cases, pre-trained audio encoders substantially improve the performance of the audio captioning system. In the language decoder, although a lot of pre-trained Transformer-based language models have been released in recent years, most of those models cannot be directly used as the language decoder, since the decoder needs to interact with audio features from the encoder via a cross-attention module. Koizumi et al. \cite{koizumi2020ac_gpt2} utilize GPT-2 \cite{radford2019gpt2} to get word embeddings. Gontier et al. \cite{Gontier2021ac_bart} fine-tune BART \cite{lewis2020bart} conditioned on the pre-trained audio embeddings and tags to generate captions and achieve state-of-the-art performance. To leverage pre-trained BERT \cite{devlin2019bert}, Liu et al. \cite{liu2022leveraging} investigate the addition of cross-attention layers with randomly initialized weights in the pre-trained BERT models as the decoder and demonstrate the efficacy of the pre-trained BERT models for audio captioning. 

In summary, pre-trained audio encoders have proved to be effective to get robust audio features and overcome the data scarcity problem, while how to incorporate existing large pre-trained language models into an audio captioning system still needs further investigation.

\begin{table*}[!ht]
\caption{An overview of published methods for audio captioning.}
\centering
\resizebox{\textwidth}{!}{
\begin{tabular}{cccccc}
\hline
 Reference & Year & Audio Encoder & Text Decoder & Key aspects \\
\hline
Drossos et al. \cite{drossos2017ac_1} & 2017 & RNN & RNN & Attention \\
Wu et al. \cite{wu2019audiocaption_hospital}& 2019 & RNN & RNN & N$\backslash$A \\
Xu et al. \cite{xu2021audiocaption_car} & 2019 & RNN & RNN & Sentence similarity loss \\
Ikawa et al. \cite{Ikawa2019ac_spec} & 2019 & RNN & RNN & `Specificity' term \\
Kim et al. \cite{kim2019audiocaps} & 2019 & CNN(VGGish)+RNN & RNN & Multi-scale features, Semantic attention \\
Nguyen et al. \cite{Nguyen2020temporalsub} & 2020 & RNN & RNN & Temporal subsampling \\ 
Cakir et al. \cite{Cakir2020ac_multitask} & 2020 & RNN & RNN & Multi-task learning (keywords) \\
Perez-Castanos et al. \cite{Perez-Castanos2020ac_gamma} & 2020 & CNN & RNN & Attention \\
Chen et al. \cite{chen2020ac_CNN} & 2020 & CNN & Transformer & Pre-trained encoder \\
Xu et al. \cite{xu2020crnn_sjtu} & 2020 & CRNN & RNN & Reinforcement learning \\
Takeuchi et al. \cite{Takeuchi2020ntt_first} & 2020 & CNN+RNN & RNN & Keywords, sentence length estimation \\
Tran et al. \cite{tran2020wavetransformer} & 2020 & CNN & Transformer & 1-D and 2-D CNN \\
Eren et al. \cite{eren2020acfake} & 2020 & CNN(PANNs)+RNN & RNN & Keywords \\
Koizumi et al. \cite{koizumi2020keywords} & 2020 & CNN(VGGish)+Transformer & Transformer & Keywords \\
Koizumi et al. \cite{koizumi2020ac_gpt2} & 2020 & CNN(VGGish) & GPT-2+Transformer & GPT-2, similar captions retrieval \\
Xu et a. \cite{xu2021invest_cnn_crnn} & 2021 & CNN$\backslash$CRNN & RNN & Attention, transfer learning \\
Mei et al. \cite{Mei2021ac_trans} & 2021 & CNN(PANNs) & Transformer & Transfer learning, reinforcement learning \\
Mei et al. \cite{Mei2021ACT} & 2021 & Transformer & Transformer & Full transformer network \\
Han et al. \cite{Han2021netease} & 2021 & CNN(PANNs) & Transformer & Weakly-supervised pre-training, keywords \\
Ye et al. \cite{Ye2021peking} & 2021 & CNN(PANNs) & RNN & Keywords, attention \\
Gontier et al. \cite{Gontier2021ac_bart} & 2021 & CNN(VGGish) & BART & YAMNet tags, BART \\
Narisetty et al. \cite{Narisetty2021ac_conformer} & 2021 &CNN(PANNs)+Conformer & Transformer+RNN & ASR techniques \\
Liu et al. \cite{Liu2021cl4ac} & 2021 & CNN(PANNs) & Transformer & Contrastive learning \\
Won et al. \cite{Won2021ac} & 2021 & CNN(PANNs) & Transformer & Transfer learning \\
Berg et al. \cite{Berg2021continual} & 2021 & CNN & Transformer & Continual learning \\
Weck et al. \cite{Weck2021ac_offtheshelf} & 2021 & CNN(VGGish,YAMNet,OpenL3,COALA) & Transformer & Transfer learning \\
Mei et al. \cite{mei2021diverse} & 2021 & CNN(PANNs) & Transformer & GAN, diversity \\
Xiao et al. \cite{xiao2022local} & 2022 & CNN & Transformer & Attention-free Transformer \\
Liu et al. \cite{liu2022leveraging} & 2022 & CNN(PANNs) & BERT & Transfer learning, BERT \\
Chen et al. \cite{chen2022contrastive} & 2022 & CNN & Transformer & Transfer learning, contrastive learning \\
Koh et al. \cite{koh2022ac_latent} & 2022 & CNN(PANNs)+Transformer & Transformer & Transfer learning, regularization\\
Narisetty et al. \cite{narisetty2022ac_asr} & 2022 & Transformer & Transformer & Joint modelling of ASR and AAC \\
\hline
\label{tab:methods_statistics}
\end{tabular}}
\end{table*}

\subsection{Other approaches}
Contrastive learning has been a popular training method in CV tasks \cite{chen2020ntxent, He2020moco}, Liu et al. \cite{Liu2021cl4ac} and Chen et al. \cite{chen2022contrastive} both investigated using contrastive training to learn better alignment between audio and text. Specifically, an audio clip and its paired caption are regarded as a positive pair while audio clips with other non-paired captions are regarded as negative pairs. A contrastive loss is combined with the original cross-entropy loss to encourage the model to maximize the similarity of the embeddings from the positive pairs while minimizing the similarity of the embeddings from the negative pairs.
Koh et al. \cite{koh2022ac_latent} also proposed an auxiliary objective aiming at maximizing the similarity between latent space formed by audio features obtained by the encoder and the latent space formed by text features obtained by the decoder.
Berg et al. \cite{Berg2021continual} presented a continual learning approach for continuously adapting an audio captioning method to new unseen general audio signals without forgetting learned information. Mei et al. \cite{mei2021diverse} argued that an audio captioning system should have the ability to generate diverse captions for a given audio clip or across similar audio clips like human beings. They proposed an adversarial training framework based on generative adversarial network (GAN) \cite{yu2017seqgan} to encourage the diversity of audio captioning systems. In addition, because ASR and AAC share similarities in translating sound into natural language, Narisetty et al. \cite{narisetty2022ac_asr} proposed approaches for end-to-end joint modeling of speech recognition and audio captioning tasks. 

A brief overview of the published audio captioning methods is shown in Table~\ref{tab:methods_statistics}, which contains the type of deep neural networks used to encode audio information, the language models used to generate captions, and the key aspects of these methods in the final column.

\section{Evaluation metrics}
\label{sec:eval_metrics}
\noindent Evaluating audio captions is a challenging and subjective task, because an audio clip can correspond to several correct captions that may use different words, grammar, and/or describe different parts of the audio clip. Existing works adopt the same evaluation metrics used in image captioning, where most of these metrics are borrowed from NLP tasks such as machine translation and summarization, and the remaining are designed specifically for image captioning. The automatic evaluation metrics compare the machine-generated captions with human-annotated references where the number of references for each audio clip may vary across different datasets. The number of parallel references will influence the evaluation results. Generally, more reference captions are favored to reduce the evaluation bias \cite{zhu2020consensus}. In this section, we first introduce the conventional rule-based metrics, and then discuss some novel model-based metrics.

\subsection{Conventional evaluation metrics}
\noindent \textbf{BLEU}. BLEU (BiLingual Evaluation Understudy) \cite{papineni2002bleu} is originally designed to measure the quality of machine-generated sentences for machine translation systems. BLEU calculates modified $n$-gram precision for $n$ up to four: the counts for $n$-grams in candidate sentence is first collected and clipped by their corresponding maximum count in references, the clipped counts are then summed and divided by the total number of candidate $n$-grams, where the $n$-gram is a window consisting of $n$ consecutive words. The modified $n$-gram precisions are averaged with uniform weights to account for both adequacy and fluency, where $1$-gram matches account for adequacy while longer $n$-gram matches account for fluency. As precision tends to give short sentences higher scores, a brevity penalty is introduced to penalize short sentences.

\noindent \textbf{ROUGE}. ROUGE (Recall-Oriented Understudy for Gisting Evaluation) \cite{lin2004rouge} includes a set of metrics proposed to measure the quality of a machine-generated summary \cite{lin2004rouge}. ROUGE-L which is widely used in image and audio captioning is based on the matching of the longest common subsequence between a candidate and a reference caption. ROUGE-L first counts the length of the longest common subsequence between a candidate and a reference, which is then divided by the total lengths of the candidate and reference to get precision and recall respectively. An F-measure combining precision and recall is then calculated as the score of ROUGE-L, which favors recall more. 

\noindent \textbf{METEOR}. METEOR (Metric for Evaluation of Translation with Explicit ORdering) \cite{banerjee2005meteor} is also a metric to evaluate machine translation systems. METEOR calculates unigram precision and unigram recall based on an explicit word-to-word matching in terms of their surface forms, stemmed forms, and meanings between a candidate and one or more references. An F-mean placing most of the weight on recall is then computed. To take into account longer matches, unigrams that are in adjacent positions in candidate and references are grouped into chunks, a penalty based on chunks is introduced and combined with F-mean to give the final METEOR score. 

\noindent \textbf{CIDEr}. CIDEr (Consensus-based Image Description Evaluation) \cite{vedantam2015cider} is an automatic consensus metric for evaluating image description quality. CIDEr also represents sentences using $n$-grams presented in them, where each $n$-gram is weighted by the term frequency inverse document frequency (TF-IDF) weights because $n$-grams that commonly occur in a dataset are likely to be non-informative. CIDEr computes the cosine similarity of weighted $n$-grams between candidate and references, which accounts for both precision and recall. Similar to BLEU, CIDEr considers higher order $n$-grams (up to four) to capture grammatical properties and richer semantics. 

\noindent \textbf{SPICE}. SPICE (Semantic Propositional Image Caption Evaluation) \cite{anderson2016spice} is an image captioning evaluation metric based on semantic content matching. SPICE parses both candidate and references into scene graphs in which the objects, attributes and relations are encoded. An F-score is then calculated based on the matching of tuples extracted from the candidate and reference scene graphs. SPICE ignores the properties of grammar and fluency of sentences but only focuses on semantic matching.

\noindent \textbf{SPIDEr}. SPIDEr \cite{liu2017spider} is proposed for evaluating image captions and used as the official ranking metric in the automatic audio captioning task in DCASE Challenge. SPIDEr is the average of SPICE and CIDEr: the SPICE score ensures captions are semantically faithful to the content, while CIDEr score ensures captions are syntactically fluent.

\subsection{Model-based metrics}
\noindent \textbf{BERTScore}. BERTScore \cite{Zhang2020BERTScore} is an evaluation metric for text generation tasks. Unlike conventional metrics which almost all rely on surface-form similarity, BERTScore utilizes pre-trained BERT \cite{devlin2019bert} contextual embeddings that can capture semantic similarity, distant dependencies and ordering. After getting contextual embedding of each token through BERT, BERTScore measures the similarity of each token between candidate and references through cosine similarity where each token is matched to the most similar token in the other sentences. The matched token pairs are used to calculate a precision, recall and an F1 measure. Importance weighting with inverse document frequency is also introduced to weight rare words more.

\noindent \textbf{SentenceBERT}. SentenceBERT \cite{reimers2019sentenceBERT} is not essentially an evaluation metric but a modification of the pre-trained BERT model \cite{devlin2019bert}. The SentenceBERT model can be used to obtain fixed-sized sentence embeddings for input captions. The sentence embeddings are then used to calculate a similarity score, such as cosine similarity, Euclidean distance or other similarities, between candidate and reference captions. Compared with BERTScore that measures the similarity in token level, SentenceBERT can be used for audio captioning for similarity comparison in sentence level.

\noindent \textbf{FENSE}. FENSE (Fluency ENhanced Sentence-bert Evaluation) \cite{zhou2021fense} is a model-based evaluation metric specifically proposed for audio captioning. FENSE utilizes the Sentence-BERT to derive sentence embeddings for candidate and reference captions, and calculates its average cosine similarity score. To capture grammar issues like repeated words or phrases and incomplete sentences, FENSE uses a separate pre-trained error detector to penalize the Sentence-BERT scores when fluency issues are detected. 

In summary, conventional rule-based metrics are widely used to evaluate the performance of audio captioning systems. Most of these metrics focus on $n$-gram or sub-sequence based matching between the generated and reference captions. CIDEr and SPICE, proposed for image captioning, have shown better correlation with human judgements in captioning tasks than those borrowed from NLP tasks \cite{vedantam2015cider, liu2017spider}. However, some authors have shown that these metrics still cannot resemble human judgment well \cite{novikova2017metrics, zhou2021fense}. Model-based metrics have received increasing attention and shown better correlation with human judgements in NLP tasks, but they have not been widely used in captioning tasks so far. We introduce them here to encourage research effort for developing novel metrics for audio captioning.

\begin{table*}[!ht]
\caption{Performances of some surveyed audio captioning methods on two main datasets.  Scores are taken from the respective papers. Only single model performance is considered. Compared to Clotho v1, Clotho v2 introduces new audio clips into the training set and a new validation set, while retaining the same evaluation set. Some methods merge the new validation set into the training set, these methods are still evaluated using the same evaluation set. We report these results separately. Highest scores for each split are shown in bold.}
\begin{tabular}{ccccccccc}
\hline
 Dataset & Method & Year & BLEU$_{1}$ & BLEU$_{2}$ & METEOR & CIDEr & SPICE & SPIDEr \\
\hline
 \multirow{7}{*}{AudioCaps} & Kim et al. \cite{kim2019audiocaps} & 2019 & 0.614 & 0.446 & 0.203 & 0.593 & 0.144 & 0.369 \\
 & Koizumi et al. \cite{koizumi2020ac_gpt2} & 2020 & 0.638 & 0.458 & 0.199 & 0.603 & 0.139 & 0.371 \\
 & Eren et al. \cite{eren2020acfake} & 2020 & \textbf{0.710} & 0.490 & \textbf{0.290} & 0.750 & - & - \\
 & Xu et al. \cite{xu2021invest_cnn_crnn} & 2021 & 0.655 & 0.476 & 0.229 & 0.660 & 0.168 & 0.414 \\ 
 & Mei et al. \cite{Mei2021ACT} & 2021 & 0.647 & 0.488 & 0.222 & 0.679 & 0.160 & 0.420 \\
 & Gontier  et al. \cite{Gontier2021ac_bart} & 2021 & 0.699 & \textbf{0.523} & 0.241 & \textbf{0.753} & \textbf{0.176} & \textbf{0.465} \\
 & Liu et al. \cite{liu2022leveraging} & 2022 & 0.671 & 0.498 & 0.232 & 0.667 & 0.172 & 0.420 \\
\hline
 \multirow{12}{*}{Clotho v1} & Drossos et al. \cite{drossos2020clotho} & 2019 & 0.420 & 0.140 & 0.090 & 0.100 & - & - \\
 & Cakir et al. \cite{Cakir2020ac_multitask} & 2020 & 0.409 & 0.156 & 0.088 & 0.107 & 0.040 & 0.074 \\
 & Nguyen et al. \cite{Nguyen2020temporalsub}& 2020 & 0.417 & 0.154 & 0.089 & 0.093 & 0.040 & 0.067\\
 & Perez-Castanos \cite{Perez-Castanos2020ac_listen} & 2020 & 0.469 & 0.265 & 0.136 & 0.214 & 0.086 & 0.150 \\
 & Tran et al. \cite{tran2020wavetransformer} & 2020 & 0.489 & 0.303 & 0.143 & 0.268 & 0.095 & 0.182 \\
 & Takeuchi et al. \cite{Takeuchi2020ntt_first} & 2020 & 0.512 & 0.325 & 0.145 & 0.290 & 0.089 & 0.190 \\
 & Koizumi et al. \cite{koizumi2020keywords} & 2020 & 0.521 & 0.309 & 0.149 & 0.258 & 0.097 & 0.178 \\
 & Chen et al. \cite{chen2020ac_CNN} & 2020 & 0.534 & 0.343 & 0.160 & 0.346 & 0.108 & 0.227 \\
 & Xu et al. \cite{xu2020crnn_sjtu} & 2020 & 0.561 & 0.341 & 0.162 & 0.338 & 0.108 & 0.223 \\
 & Eren et al. \cite{eren2020acfake} & 2020 & \textbf{0.590} & 0.350 & \textbf{0.220} & 0.280 & - & - \\ 
 & Xu et al. \cite{xu2021invest_cnn_crnn} & 2021 & 0.556 & 0.363 & 0.169 & 0.377 & \textbf{0.115} & \textbf{0.246} \\
 & Koh et al. \cite{koh2022ac_latent} & 2022 & 0.551 & \textbf{0.369} & 0.165 & \textbf{0.380} & 0.111 & \textbf{0.246} \\
 \hline
 \multirow{4}{*}{Clotho v2} & Narisetty et al. \cite{Narisetty2021ac_conformer} & 2021 & 0.536 & 0.341 & 0.160 & 0.346 & 0.108 & 0.227 \\
 & Won et al. \cite{Won2021ac} & 2021 & 0.564 & 0.376 & \textbf{0.177} & 0.441 & \textbf{0.128} & 0.285 \\
 & Ye et al. \cite{Ye2021peking} & 2021 & 0.577 & - & 0.174 & 0.419 & 0.119 & 0.269 \\
 & Han et al. \cite{Han2021netease} & 2021 & \textbf{0.585} & \textbf{0.392} & \textbf{0.177} & \textbf{0.474} & \textbf{0.130} & \textbf{0.302} \\
 \hline
 \multirow{5}{*}{Clotho v2 + val set}
 & Narisetty et al.\cite{Narisetty2021ac_conformer} & 2021 & 0.541 & 0.346 & 0.161 & 0.362 & 0.110 & 0.236 \\
 & Liu et al. \cite{Liu2021cl4ac} & 2021 & 0.553 & 0.349 & 0.168 & 0.368 & 0.115 & 0.242 \\
 & Mei et al. \cite{Mei2021ac_trans} & 2021 & 0.561 & 0.374 & 0.171 & 0.426 & \textbf{0.124} & 0.275\\
 & Chen et al. \cite{chen2022contrastive} & 2022 & 0.572 & 0.379 & 0.171 & 0.407 & 0.119 & 0.263 \\
 & Xiao et al. \cite{xiao2022local} & 2022 & \textbf{0.578} & \textbf{0.387} & \textbf{0.177} & \textbf{0.434} & 0.122 & \textbf{0.278} \\
\hline
\label{tab:methods_scores}
\end{tabular}
\end{table*}

\section{Datasets}
\label{sec:datasets}
\noindent The release of high quality audio captioning datasets has greatly promoted the development of this area. Almost all existing datasets (except one) are collections of single-sentence English captions, however, these datasets differ in many aspects such as the number of audio clips, the number of captions per audio clip, and the length of each audio clip. These different characteristics will affect the design and the performance of the audio captioning model. We describe the details of existing datasets in this section. To better understand the datasets, we then use consensus score of previously introduced metrics to evaluate these datasets. 

\subsection{Datasets description}
\noindent \textbf{AudioCaps}. AudioCaps \cite{kim2019audiocaps} is the largest audio captioning dataset so far. All the audio clips are 10-seconds long and are sourced from AudioSet, a large-scale audio event dataset \cite{audioset}. The audio clips are selected by following some selection qualifications that ensure the chosen audio clips are balanced with respect to the ground truth annotations (tags) in the original dataset and diverse in terms of content. The audio clips are annotated by crowdworkers through Amazon Mechanical Turk (AMT), annotators are provided with an audio clip with corresponding word hints and video hints, and are required to write a natural language description with provided information.  

The official release of AudioCaps contains 51k audio clips and is divided into a training set, a validation set and a test set. Each audio clip in the training set contains one corresponding human-annotated caption while those in validation set and test set contain five corresponding captions. Audio clips in AudioSet are not freely available but can be extracted from YouTube videos. It is worth noting that some audio clips might be no longer downloadable, thus the number of downloadable audio clips might be different from the official release of AudioCaps. The statistics in Table~\ref{tab:datasets_statis} are reported based on the official release version of AudioCaps.

\begin{table}[!th]
\caption{An overview of English-annotated datasets.}
\centering
\resizebox{\linewidth}{!}{
\begin{tabular}{cccccc}
\hline
 Dataset & \ \thead{\# of\\ audios} & \thead{\# of captions\\ per audio} & \thead{Audio \\ duration} & \thead{Vocab \\ size} & \thead{Avg caption\\lengths} \\
\hline
AudioCaps & \num{51308} & 1, 5 & 10 s & 5066 & 8.79 \\

Clotho & \num{5929} & 5 & 15-30 s & 4365 & 11.33 \\

MACS & \num{3930} & 2, 3 ,4, 5 & 10 s & 2776 & 9.24 \\
\hline
\label{tab:datasets_statis}
\end{tabular}}
\end{table}

\noindent \textbf{Clotho}. Clotho \cite{drossos2020clotho} is the dataset used for official ranking of the submitted systems in the task 6 (Automated Audio Captioning) of DCASE challenges in 2020 and 2021. All the audio clips are sourced from the online platform Freesound \cite{font2013freesound} and are ranging almost uniformly from 15 to 30 seconds. Annotators are employed through AMT for crowdsourcing the captions. During the annotation process, only the audio signal was provided to the annotators, with no additional information such as word or video hints (different from AudioCaps), to avoid introducing biases.

The latest Clotho v2 published a development set containing three subsets. There are \num{3839} audio clips in the training set and \num{1045} audio clips in the validation and evaluation set, respectively. Each audio clip contains five human-annotated captions, ranging from 8 to \num{20} words long. In the DCASE challenges, all three of these published sets can be used to train the models, while the final performance is evaluated using a preserved testing split by the organisers. For reporting results for conference or journal papers, performances are assessed using the published evaluation set and some authors may include the validation set into training since the validation set is added in Clotho v2. As a result, the model performance reported on Clotho may not all be on the same ground. 

\begin{table*}[!ht]
\caption{Consensus scores of English-annotated datasets.}
\centering
\resizebox{\textwidth}{!}{
\begin{tabular}{cccccccccccc}
\hline
 & \ BLEU$_1$ & BLEU$_2$ & BLEU$_3$ & BLEU$_4$ & ROUGE-L & METEOR & CIDEr & SPICE & SPIDEr & BERTScore & SentenceBERT \\
\hline
AudioCaps & 0.65 & 0.48 & 0.37 & 0.29 & 0.49 & 0.28 & 0.90 & 0.21 & 0.56 & 0.52 & 0.64 \\

Clotho & 0.65 & 0.49 & 0.38 & 0.31 & 0.50 & 0.30 & 0.86 & 0.23 & 0.54 & 0.54 & 0.53 \\

MACS & 0.49 & 0.28 & 0.16 & 0.08 & 0.32 & 0.18 & 0.21 & 0.13 & 0.17 & 0.24 & 0.52 \\
\hline
\label{tab:datasets_consensus}
\end{tabular}}
\end{table*}

\noindent \textbf{MACS}. MACS (Multi-Annotator Captioned Soundscapes) \cite{Martin2021databias} consists of audio clips from the development set of TAU Urban Acoustic Scenes 2019 dataset \cite{Mesaros2018TAU}. The audio clips are all 10-second long recorded from three acoustic scenes (airport, public square and park) and are annotated by students. The annotation process contains two stages. Given a list of ten classes and an audio clip, the annotators are first required to select the audio events presented in an audio clip from the given class list. Afterwards, the annotators are required to write a description of the audio clip. 

MACS contains \num{3930} audio clips without being split into subsets. The number of captions per audio clip varies in the dataset. Most audio clips have five corresponding human-annotated captions, while some of them may only have two, three or four. 

\noindent \textbf{AudioCaption} AudioCaption is a domain-specific Mandarin-annotated audio captioning dataset. Two scene-specific sets have been published: one for hospital scene \cite{wu2019audiocaption_hospital} and another for car scene \cite{xu2021audiocaption_car}. The hospital-scene set contains \num{3707} audio clips with three captions per clip while the car-scene set contains \num{3602} audio clips with five captions per clip. All the audio clips are annotated by native Mandarin speakers.

\subsection{Datasets evaluation}
\noindent Since all the datasets mentioned above are annotated under different protocols, they show different characteristics such as the number of captions per audio clip, caption lengths and sample variance in multi-reference captions. We believe these characteristics will influence the design and performance of audio captioning models. To better understand the datasets, we evaluate three English-annotated datasets from different aspects. Table~\ref{tab:methods_scores} reports the performance of some surveyed methods on two main datasets, AudioCaps and Clotho, for which some methods listed in Table ~\ref{tab:methods_statistics}, such as \cite{drossos2017ac_1, Ikawa2019ac_spec, xu2021audiocaption_car}, are not considered as they were not evaluated on these two datasets.
Table \ref{tab:datasets_statis} summarizes the datasets with some basic statistics. In addition, we use a consensus score \cite{zhu2020consensus} to represent the agreement among the parallel reference captions for the same audio clip, and the results are shown in Table~\ref{tab:datasets_consensus}. The consensus score $c$ among $n$ parallel reference captions $\mathcal{R}=\{r_i\}_{i=1}^n$ for an audio clip is defined as:
\begin{equation}
\label{eqn:consensus}
c=\frac{1}{n}\sum_{i=1}^n{\rm{metric}}(r_i, \mathcal{R}\backslash{r_i})
\end{equation}
where $r_i$ is the $i$-th caption and the metric can be anyone mentioned above. Since the number of references are varied among different datasets, we report the consensus score of AudioCaps using the validation and test set, Clotho using the training set and MACS using all the audio clips having five reference captions. 

As the consensus scores are computed among the human-annotated captions, they can be also regarded as upper bound human-level performance on each dataset. As can be seen from Table~\ref{tab:datasets_consensus}, the consensus scores on AudioCaps and Clotho are close to each other except that the SentenceBERT score on AudioCaps is clearly higher than that of Clotho.  Surprisingly, the consensus scores on MACS are lower than the other two datasets while only the SentenceBERT is close to them. This may reveal that the human-annotated captions in MACS are more diverse than the other two, and SentenceBERT can better capture semantic relevance between diverse captions. The consensus scores can be regarded as a measure of the dataset quality to some extent.

\section{Challenges and future directions}
\label{sec:challenges}
Many deep learning-based methods have been proposed to improve automated audio captioning systems, and this task has seen rapid progress in recent years. However, there is still a large gap between the performance of the resulting systems and human level performance. In this section, we discuss challenges remaining in this area and envisage possible future research directions.

\subsection{Data}
There are two main challenges about data for audio captioning. First, the data scarcity problem is still a main challenge. Existing datasets are limited in size. The collection of an audio captioning dataset is time consuming, and it is hard to control the quality of human-annotated captions. Han et al. \cite{Han2021netease} collect weakly labeled dataset from online available sources to pre-train the AAC model and show that more training data (even weakly-labelled) can greatly improve the system performance. This reveals that we can make use of audio clips available online with their weakly-labelled text description to learn more robust audio-text representation, such as CLIP \cite{radford2021clip} in computer vision.

Second, existing datasets usually do not cover all possible real-life scenarios, and thus, audio captioning models cannot generalize well to different contexts. Martin et al. \cite{Martin2021databias} investigate dataset bias of existing datasets from a lexical perspective. The bias problem still needs more investigation, e.g. how it will influence the model performance.

\subsection{Model and training strategies}
Existing AAC methods all follow the encoder-decoder paradigm and generate sentences in an auto-regressive manner. These two techniques have been the standard recipe for audio captioning models. Nonetheless, novel methods should be investigated in future research. For example, BERT-like architectures which fuse acoustic and textual modalities in early stage can be a replacement for the encoder-decoder paradigm, and work well in image captioning \cite{zhou2020unified, li2020oscar}. Non-auto-regressive language models could reduce the inference time by generating all words in parallel \cite{gao2019non_auto}, which might be a worthwhile research direction as it offers computational advantages, despite the fact that it under-performs the auto-regressive models in terms of captioning accuracy.

For the training strategies, the standard cross-entropy loss brings the problem of `exposure bias' and tends to generate simple and generic captions. Although reinforcement learning is introduced to solve this problem, it may adversely affect the quality of generated captions. A promising line of research is to design new objective functions or add human feedback in a reinforcement learning setting to solve these problems. In addition, it requires more investigation on how to make use of learned knowledge in large pre-trained language models to improve caption generation.

\subsection{Evaluation}
The performance of audio captioning systems is generally assessed by objective evaluation metrics, since the human evaluation can be time-consuming and expensive. As discussed in the end of Section~\ref{sec:eval_metrics}, existing objective metrics may not correlate well with human judgements \cite{Mei2021ac_trans, novikova2017metrics, zhou2021fense}, and none of them are designed specifically for audio captioning. Future work is expected to figure out to what extent the existing objective metrics correlate with human judgements, and to develop more reliable evaluation metrics.

\subsection{Diversity and stylized captions}
As argued in \cite{dai2017im_diverse}, a good captioning model should generate sentences that possess three properties: fidelity, i.e. the generated captions should reflect the audio content faithfully; naturalness, i.e. the captions should not be identified as machine-generated; diversity, i.e. the sentences should have rich and varied expressions, reflecting how different people would describe an audio clip in different ways. However, many existing approaches only consider semantic fidelity. Further research should be conducted to improve the other two properties. In addition, stylized captioning systems, which can generate outputs suitable for different audiences such as kids, could be a worthwhile research direction.

\subsection{Other potential directions}
There are also other potential directions for audio captioning. For example, temporal information of the sound events is not well used in existing works. Future work could investigate the use of information related to activities and timing information of sound events to generate more accurate captions. Information from other modalities could be also employed to train the audio captioning models, such as using audio-visual captioning methods \cite{tian2018attempt, iashin2020better}. In addition, audio captioning can be potentially linked with other audio-language multi-modal tasks, such as audio-text retrieval \cite{koepke2022audioretrieval, mei2022_t6b}, audio question answering \cite{fayek2020temporal}, text-based audio generation \cite{liu2021conditional} and text-based audio source separation \cite{liu2022separate}.

\section{Conclusion}
\label{sec:conclu}
Audio captioning is a fast developing task involving both audio signal processing and natural language processing. In this paper, we have reviewed published audio captioning methods from the perspective of audio encoding and text decoding. We discussed auxiliary information employed to guide the caption generation, and training strategies adopted in the literature. In addition, main evaluation metrics and datasets are reviewed. We briefly outlined challenges and potential research directions in this area. We hope this survey can serve as a comprehensive introduction to audio captioning and encourage novel ideas for future research. 


\begin{backmatter}

\section*{Acknowledgements}
The authors acknowledge the insightful comments provided
by the Associate Editor and the anonymous reviewers, which
have added much to the clarity of the paper.
For the purpose of open access, the authors have applied a creative commons attribution (CC BY) licence to any author accepted manuscript version arising.

\section*{Funding}
This work is partly supported by a Newton Institutional Links Award from the British Council, titled ``Automated Captioning of Image and Audio for Visually and Hearing Impaired" (Grant number 623805725), a grant from the Engineering and Physical Sciences Research Council (EPSRC) with number EP/T019751/1, and a Research Scholarship from the China Scholarship Council (CSC) No.202006470010.  

\section*{Abbreviations}
AT: Audio tagging, SED: Sound event detection, ASC: Acoustic scene classification, AAC: Automated audio captioning, NLP: Natural language processing, ASR: Automatic speech recognition, RNN: Recurrent neural network, CNN: Convolutional neural network, STFT: Short time Fourier transformer, MFCCs: Mel-frequency cepstral coefficients, DCT: Discrete cosine transform, GRU: Gated recurrent unit, LSTM: Long-short term memory, CV: Computer vision, CRNN: Convolutional recurrent neural network, MLP: Multi-layer perceptron, CE: Cross-entropy, MLE: Maximum likelihood estimation, GAN: Generative adversarial network.

\section*{Availability of data and materials}
The datasets analysed during this article are available on the internet.

\section*{Authors' contributions}
XM was a major contributor in writing the manuscript. XL summarized challenges and future work. MDP and WW substantially revised the manuscript. All authors read and approved the final manuscript.

\section*{Competing interests}
WW is an editorial board member of EURASIP Journal on Audio Speech and Music Processing and also a guest editor of the special issue "Recent Advances in Computational Sound Scene Analysis", other authors declare that they have no competing interests. 

\section*{Authors' information}
The authors are with Centre for Vision Speech and Signal Processing (CVSSP), Department of Electrical and Electronic Engineering, Faculty of Engineering and Physical Sciences, University of Surrey, Guildford, GU2 7XH, UK. E-mail: [x.mei, xubo.liu, m.plumbley, w.wang]@surrey.ac.uk.


\bibliographystyle{IEEEtran} 
\bibliography{bmc_article}      

\end{backmatter}
\end{document}